%% file: ms.tex
\documentclass[12pt, preprint]{aastex}
\newcommand{\ms}{\mbox{m s$^{-1}~$}}

\newcommand{\kms}{\mbox{km s$^{-1}~$}}

\newcommand{\rjup}{R$_{\rm JUP}~$}

\newcommand{\teff}{\mbox{$T_{\rm eff}$}}
\newcommand{\teq}{\mbox{$T_{\rm eq}$}}
\newcommand{\tint}{\mbox{$T_{\rm int}$}}
\newcommand{\etal}{\mbox{\rm et al.~}}

\newcommand{\lessim}{\lower.5ex\hbox{$\; \buildrel < \over \sim \;$}}

\newcommand{\simgt}{\lower.5ex\hbox{$\; \buildrel > \over \sim \;$}}
\newcommand{\simlt}{\lower.5ex\hbox{$\; \buildrel < \over \sim \;$}}

\received{}
\accepted{}

\slugcomment{Submitted to PASP; Revised July 25, 2006}

\shortauthors{Bozorgnia \etal}
\shorttitle{Search for an Atmosphere in HD149026}
\begin{document}

\title{Search for an Atmospheric Signature\\
     of the Transiting Exoplanet HD 149026b\altaffilmark{1}}
\author{Nassim Bozorgnia\altaffilmark{2},
Jonathan J. Fortney\altaffilmark{3,4,5}
Debra A. Fischer\altaffilmark{2},
Chris McCarthy\altaffilmark{2},
Geoffrey W. Marcy\altaffilmark{2,6}}

\altaffiltext{1}{Based on observations obtained at the
W. M. Keck Observatory, which is operated by the University of  
California
and the California Institute of Technology. Keck time has been
granted by NASA.}
\altaffiltext{2}{Department of Physics \& Astronomy, San Francisco State
University, San Francisco, CA  94132; fischer@stars.sfsu.edu}
\altaffiltext{3}{Space Science and Astrobiology Division, NASA Ames  
Research
Center, MS 245-3, Moffett Field, CA 94035; jfortney@arc.nasa.gov}
\altaffiltext{4}{Spitzer Fellow}
\altaffiltext{5}{SETI Institute, 515 North Whisman Road, Mountain  
View, CA 94043}
\altaffiltext{6}{Department of Astronomy, University of California,
Berkeley, CA USA 94720}

\begin{abstract}

HD149026b is a short-period, Saturn-mass planet that transits a metal- 
rich
star.  The planet radius, determined by photometry, is remarkably small
compared to other known transiting planets,
with a heavy element core that apparently comprises $\sim 70$\% of  
the total
planet mass.  Time series spectra were obtained at Keck before and
during transit in order to model the Rossiter-McLaughlin effect.
Here, we make use of these observations to carry out a differential
comparison of spectra obtained in and out-of-transit
to search for signatures of neutral atomic lithium and potassium from  
the planet atmosphere.  No signal was detected at the 2\% level; we  
therefore place upper limits on the column density of these atoms.


\end{abstract}


\section{Introduction}

Extrasolar planets that transit their host stars offer a
unique opportunity to search for constituents in the planet atmosphere.
As stellar flux passes through the planet atmosphere, additional
absorption features will be imprinted on the (otherwise pure) stellar
spectrum (Seager \& Sasselov 2000, Brown 2001, Hubbard \etal 2001).   
This offers a possibility
for detection of atmospheric transmission features
when working differentially with in-transit and out-of-transit spectra. 
We have used this fact to search for additional absorption in alkali lines  
from the atmosphere of HD 149026b during transit.

Transiting planets that reside close
to their host stars are strongly irradiated and have atmospheres
with temperatures (but not gravities) similar to those of late type
L dwarfs.  In the cool atmosphere of sub-stellar objects
like brown dwarfs and giant exoplanets, the more refractory metals
condense and rain out-of-the atmosphere leaving the less refractory
alkali metals in atomic form (Burrows \& Sharp 1999; Lodders 1999).
Thus, the strongest absorption features are expected for alkali
metals such as sodium and potassium
(Seager \& Sasselov 2000, Sudarsky, Burrows \& Pinto 2000).

Since the discovery of HD 209458b
(Charbonneau et al. 2000; Henry et al. 2000),
this star has been the subject of many ground-based and
space-based studies.  HD 209458 is a bright star with
$V=7.65$, permitting spectroscopy with high resolution and high S/N.
The transiting planet has an anomalously large radius of 1.3 \rjup,  
making
it an excellent candidate to search for atmospheric signatures. In  
contrast,
searching for atmospheric features in other transiting exoplanets
may be more difficult, because the planet to star radius is generally
smaller and the host stars have a lower intrinsic brightness.   
However, if
other exoplanets have tenuous, but extended atmospheres, then it may be
possible to identify atmospheric features from the planet.

The most successful searches for exoplanet atmospheres have been carried
out from space where telluric contamination is avoided and
higher S/N has enabled more sensitive searches.  Based on
observations with the \emph{Hubble Space Telescope} (HST),
Charbonneau \etal (2002) detected absorption from the sodium resonance
doublet at 5893 \AA\ in the dense lower atmosphere of HD 209458b, and
Vidal-Madjar \etal (2003) detected an escaping, trailing extended
exosphere of hydrogen.  In contrast, it is more challenging
to search for exoplanet atmospheres with ground-based telescopes.
Bundy \& Marcy (2000) carried out a differential analysis of spectra
for HD 209458 obtained in and
out-of-transit at the Lick and Keck Observatories.  They also studied  
the
spectra of 51 Peg, which is not a transiting planet.  Because of
contamination from ${\rm H_2O}$ and ${\rm O_2}$ features, they were only
able to place weak upper limits on Na I and K I line depth variations.
Deming et al. (2005) searched for carbon monoxide
absorption features in the transmission spectrum of HD 209458b
at $\sim2.3$ $\micron$ using NIRSPEC on Keck II telescope.  Their  
analysis
required 1077 high-resolution spectra to achieve
comparable sensitivity to the HST sodium observations.
No changes in the CO features in and out-of-transit were measured,
implying the presence of high clouds in the planet atmosphere.

\section{HD 149026: A Star with a Transiting Planet}

HD 149026 is a metal-rich G0 IV subgiant with V=8.15 and an absolute  
visual
magnitude, $M_V=3.66$. The planet's mass is similar to Saturn in our  
solar
system with a radius of $0.725$ \rjup (Sato \etal 2005)
determined from the observed photometric transit depth.
Interior models discussed by Sato \etal (2005) imply a
planet core of about 70 Earth masses.  HD 149026b orbits with a
semimajor axis of only $a=0.045$ AU, and Sato et al. (2005) report an
effective temperature in the planet atmosphere of 1540 K.

\subsection{Expected Atmospheric Features}

The equilibrium temperature of an exoplanet atmosphere at the substellar
point is a function of the effective stellar temperature, $T_*$, the
stellar radius,
$R_*$, and the planet-star separation, $a$. These factors are modulated
by the planet Bond albedo, $A_B$, and by a geometrical factor, $f$.
If a thermal gradient is established from heating on the tidally  
locked day
side  of the planet, then a wind will circulate energy to the night side
of the planet. Depending on the efficiency of this process, $f$ is 1 for
isotropic emission from both sides of the planet, and $f$ scales to 2
for emission from the day side only.
\begin{equation}
T_{eq}=T_*(R_*/2a)^{1/2}[f(1-A_B)]^{1/4}
\end{equation}
Adopting $T_*$ and $R_*$ from Sato \etal (2005), and assuming
isotropic emission and a Bond albedo, $A_B = 0.3$, we calculate
$T_{eq}=1593$ K.

Fortney \etal (2006) define the planet effective temperature as:

\begin{equation}
T_{eff}^{4}=T_{eq}^{4}+T_{int}^{4}
\end{equation}

\noindent
where \teq is the equilibrium temperature of the planet in the presence of stellar radiation,
and \tint is the effective intrinsic temperature of the planet in isolation. 
These authors note that for strongly
irradiated planets, the intrinsic isolation temperature is negligible.
They consider atmospheric models for HD 149026b with
a range of metallicities: [M/H] = 0., 0.5, 1.0.
For a cloud-free model with [M/H]=0.5 and assuming isotropic
radiation, Fortney \etal (2006) derive an effective
temperature for the planet atmosphere of 1734 K, implying $A_B < 0.1 
$.  If radiation is
not isotropic, but radiates from the day side only (2$\pi$ steradians),
this same atmospheric model yields \teff $ = 2148$ K, and TiO and VO
features appear at low pressures (Fortney \etal 2006).

The temperature of HD 149026b is therefore likely to be similar to
the temperature of L/T dwarfs (Kirkpatrick 2005).
The brown dwarf spectral sequence is a temperature
sequence over part of the range, except from mid-/late-L to mid-T  
where the
behavior of clouds become a dominant effect (Kirkpatrick 2005).
The effective temperature estimates for HD 149026b ranging from \teff= 
$1540 -2148$ K
correspond to the brown dwarf spectral type range from early-L to mid-T.

Based on the effective temperature range of the planet, observable
spectral features in the atmosphere can be predicted.  The spectra
of L dwarfs, T dwarfs and irradiated giant planets have
dominant absorption lines from neutral alkali metals like Na I, K I,
Rb I, Cs I and Li I.  Early L-type dwarfs
have prominent molecular lines, TiO and VO bands,
hydride bands FeH, CaH, and CrH in their atmospheres.
By mid to late L-types, the hydrides and the ground state neutral
lines of Na I and K I have grown in strength and are the dominant
absorbers, while the TiO and VO have weakened.
Another important feature in the spectra of late L-type
and mid T-type brown dwarfs is H${_2}$O; the neutral alkali
lines are still strong, and the hydrides are less prominent.  In
late T-type brown dwarfs, the two major lines of Na I and K I have grown
very wide (Kirkpatrick 2005).  Because the rainout of condensates
clear the atmosphere of most of its metals, the less
refractory neutral alkali metals become important in the
atmospheres of strongly irradiated, close-in ($\sim0.05$ AU)
exoplanets with \teff\ from 800 to
2000 K (Burrows \& Volobuyev 2002; Sudarsky \etal 2003).
Sodium and potassium can be particularly important both in
brown dwarfs and in the hotter ($\geq\ 1400$ K) of the giant
exoplanet classes.  Other alkali metals such as lithium should
also be present in giant exoplanets, but with lower
abundances (Sudarsky et al. 2003).
Fortney et al. (2006) find that model atmospheres of
HD 149026b with [M/H]=0.5 should show strong absorption by Na and K
in the optical spectra.  As the \emph{majority} of the planet's mass  
must be made up of metals, the metallicity of the atmosphere of HD  
149026b could be quite high.


\subsection{Observations}

A total of 30 spectra of HD 149026 were obtained at the Keck  
telescope, using
the HIRES echelle spectrometer (Vogt et al. 1994).  Eighteen of
these spectra were obtained while the planet was transiting the host  
star.
The spectra have a resolution, $R\sim$55000, and span a wavelength
range of 3500-8000 \AA\ with typical signal-to-noise ratio of 250.
All of these observations were obtained in order to derive velocity
information about the star and to model the Rossiter-McLaughlin
effect (Wolf \etal 2006). The Rossiter-McLaughlin effect
is an apparent deviation in the radial velocity from the Keplerian orbit. 
This occurs when an extrasolar planet transits across the face of its 
host star, first obscuring the approaching limb of the 
rotating star and then the receding limb of the star. The resulting 
line profile variations at the blue and red edges of the spectral 
lines is spuriously interpreted as a Doppler shift in our analysis. 

In order to carry out Doppler analysis,
an iodine cell is positioned in front of the spectrometer slit
to imprint narrow reference iodine lines.  These iodine lines
span the wavelength range
4800-6200 \AA\, and have a depth of about 20\%.  The iodine lines are
used to measure wavelength shifts (velocities) and to derive
the instrumental point spread function; however, they
interfere with our ability to measure small changes in line
absorption strengths.  Unfortunately the Na I doublet is contaminated  
by the
presence of iodine and could not be included in our analysis.

Redward of 6000 \AA, telluric lines begin to appear in the
stellar spectrum.  Differences in barycentric velocities at the times
of observations
create significant shifts in the stellar spectrum relative to
the telluric features. Telluric lines that shift in and out of the
spectral lines in our analysis are problematic because they will
introduce a spurious signal. To avoid this problem, we
cross-correlated a telluric spectrum with each stellar spectrum and
masked the regions of the spectra containing telluric lines.
For those cases where the telluric lines fall
on a stellar line, we masked out the entire spectral line to avoid
introducing large errors.  Doppler shifts arising from dynamical orbital
motion correspond to subpixel shifts and can be ignored.

\section{Analysis}

Our goal was to search for constituents of the illuminated planet
atmosphere during transit.
In order to carry out the strongest possible analysis, we worked
differentially, comparing spectra obtained during transit with
those obtained out-of-transit.  As a first step, we identified
the in-transit and out-of-transit spectra by phase-folding the
radial velocity (RV) data with a period of
$P=2.875965 +0.000085 -0.000135$ days.  This value was calculated
by taking the mean of the values provided by Charbonneau et al. (2005),
and Wolf et al. (2006) which agree with each other within the
stated uncertainties.  The time at the center of transit,
$T_c (HJD)=2453527.87455+0.00085-0.00091$, was adopted from
Charbonneau et al. (2005).

Figure 1 represents an update to the phased radial velocity data
for HD 149026 presented in Sato et al. (2005).
The Keplerian fit is overplotted on the RV data.  Fifteen RV  
measurements
made at Keck during transit are shown as open diamonds in the plot.  
These
points were removed when fitting the Keplerian model, because they
exhibit the Rossiter-McLaughlin effect and deviate from Keplerian
velocities.  We froze the period at $P=2.87596$ days to make this fit,
and the RMS RV scatter to the fit is 3.07 \ms consistent with our  
velocity
precision.  The dashed lines in
the plot show the 215 minute transit window.

We constructed a high signal-to-noise, continuum normalized template for
HD 149026 by averaging all 12 nontransit spectra.
Each individual spectrum (both in and out-of-transit)
was normalized and cross-correlated with the template
spectrum.  As a first step, we divided each spectrum by the template,  
and
searched the full
wavelength range of 6500-8000 \AA\ by eye for any residuals relative to
the template spectrum.   We examined 9 spectral orders this way
and found no significant changes in the spectrum greater than a few
percent.  Because of contamination from telluric and iodine
lines we were not able to make use of other spectral orders.

We then carried out a more robust analysis by searching individual lines
such as potassium and lithium in detail.
We considered a wavelength bin centered on the K line at 7698 \AA.  The
template spectrum is overplotted on an in-transit spectrum in Figure 2.
The difference of the two spectra (offset from zero) and
the line core bin for the K line are also shown in the same figure.
We calculated the percent deviation at each pixel by taking
the difference between each spectrum and the template spectrum, and  
dividing
by the template spectrum.
In order to average over photon noise, we binned the percent  
difference in
wavelength, choosing bins with widths comparable to the sizes of the
absorption line cores (8 pixels or 10.4 \kms).  The percent  
deviations are
plotted with respect to orbital phase in Figure 3.

The individual spectra have signal-to-noise better than 250, so we 
expect the photon noise to be about 0.4\% in the continuum but larger
in cores of the spectral lines.  In addition to photon noise, there 
are systematic errors that arise from imperfect continuum normalization, from 
the uncertainties in vertically registering the individual spectra against
the template, from spectrum-to-spectrum variations in weak telluric 
lines and perhaps from trace iodine features.

To assess our uncertainties, we computed the standard deviation of
the percent differences for 12 {\em out-of-transit} spectra in the 
wavelength bins of interest. Since the out-of-transit spectra should 
not be contaminated by the planet atmospheres,  
this represents the typical uncertainties, including Poisson errors and 
systematic errors.  The
dashed lines in Figure 3 show the limits where the variations
would have exceeded three times the intrinsic fluctuations measured in
the out-of-transit spectra. 

A Kolmogorov-Smirnov (K-S)
test was applied to the in-transit and out-of-transit residuals
and suggests that both sets of spectra were drawn from
the same parent population.  The probability that the two samples were
drawn from different populations is only 0.36\%, implying that there
is no evidence for an excess of potassium in the irradiated planet
atmosphere.

The same analysis was applied to the Li I resonance doublet at  
6707.82 \AA.
Figure 4 shows a comparison of the template and an in-transit
spectrum; the 10.4 \kms bin used for the lithium line is also plotted
in the same figure.
The percent deviations for Li are shown in Figure 5, and the dashed
lines show the limit of 3 times the intrinsic fluctuations measured in
the out-of-transit spectra.  Again,
the K-S test shows that the in-transit spectra are statistically
indistinguishable from the out-of-transit spectra; no excess lithium
absorption is seen during transit.

We also scrutinized observations taken near ingress and egress times,
for evidence of a leading or trailing atmosphere that could enhance
absorption when the planet is not actually transiting.
To investigate this possibility, we modified our original analysis of
the potassium and lithium lines.  In this analysis we constructed a
template spectrum from only five out-of-transit spectra
which were clearly not near the ingress and egress points.
Then we repeated the analysis described above for the K and Li lines.
Figure 6 shows the percent deviations for individual observations of  
the K
line relative to the new template and plotted against orbital phase.
Figure 7 shows the percent deviations for the Li I line as a function of
orbital phase.  The dashed horizontal lines again show boundaries
that represent three times the RMS scatter in the 5 non-transiting
spectra.
The nontransit observations near ingress or egress are shown as
boundary nontransit points (open triangles) in the plot.
Again, the K-S test shows that the line depths do not vary significantly
whether the observation was obtained in-transit, near ingress or  
egress, or
out-of-transit.  Therefore, we see no line enhancement that suggest a  
leading
or trailing exosphere with a large optical depth.

\section{Detectability Simulations}

To determine the increase in optical depth in the planet atmosphere that
would be needed to produce a detectable signal, we generated synthetic
spectra with injected planetary signals of different strengths.  We  
scaled
the K line and the Li line in 1\% steps for each of our {\em in-transit}
spectra. For each of these pseudo optical depth scaling factors, we
applied the two-sided K-S test to find the probability that the  
simulated
transit data
were drawn from the same distribution as our out-of-transit template.
For our analysis, we used the template constructed from five
observations that were not close to ingress or egress.
The pseudo optical depth scaling factors for the K line and the
resulting K-S statistic,
as well as the detection probabilities are listed in Table 1.
Small values of the K-S statistic show that the cumulative distribution
function of our scaled up in-transit spectra is significantly different
from the out-of-transit template, and therefore yield a high probability
detection. Alternatively, the detection probability is one minus the
K-S statistic.  A signal with 1.03 pseudo optical depth scaling factor
would have been detected at the 95.16\% confidence level.

Figure 8 shows a comparison of in-transit data with a 1.02 pseudo  
optical
depth scaling for the K line.  This synthetic data set represents a
transit by a planet which imprints a
spectral signature 2\% stronger than the signal already present in the
stellar spectrum.  The top figure shows
an overplot of the template and synthetic spectra.  We
subtracted the template from the synthetic spectrum and offset the
differences from zero for visibility on the same scale.
The bottom plot in Figure 8 shows the percent deviations
for the out-of-transit and synthetically
enhanced in-transit data, relative to the template.
The K-S test compares these two populations (out-of-transit data
and synthetic in-transit data) to calculate the detection probability
listed in Table 1.

Figure 9 shows the synthetic signal corresponding to a 1.05
pseudo optical depth scaling factor for the K line.
As with Figure 8, the simulated change of 5\% in the transit
spectra is overplotted on the template in the top plot.
Figure 9 (bottom plot) shows the percent difference
 in the 10.4 \kms bin.
A K-S test finds a detection probability of 98.8\% for this case.
Finally, Figure 10 shows the fake injected signal for a pseudo  
optical depth
factor of 1.08 at the place of the K line.  Clearly, an enhancement in
line depth of 8\% could even be discovered by simple inspection.

We carried out a similar scaling for the Li line at the
10.4 \kms bin at 6707.8 \AA\, and simulated a line with deeper strength.
Table 2 shows different values of the pseudo optical depth scaling
factor, their corresponding K-S significance level, and the
confidence level for detecting the signal.
A pseudo optical depth scaling factor of 1.02 results in a 84\%
confidence level for detecting lithium, and for a 1.08 scaling
factor, the confidence level for detecting the signal is 98.8\%.

\section{Discussion}
The detection of absorption features in the lower atmosphere of a  
transiting giant planet requires outstanding precision.  Charbonneau  
\etal (2002), who detected neutral atomic sodium, and Deming \etal  
(2005), who placed upper limits on CO absorption, obtained a  
precision 100 times better for HD209458 than we have in our study.   
They were looking for 0.02\% effects.  These studies made use of the  
\emph{Hubble Space Telescope} and over 1000 ground-based spectra,  
respectively.  However, The detection of species in an extended  
exosphere need not necessarily require this same precision, because  
the atmosphere's cross-sectional area is significantly larger (Vidal- 
Madjar \etal 2003).

If one thinks of our 2\% precision as the smallest cross-sectional  
atmospheric area that we could observe, this corresponds to 2.1 $R_ 
{\rm J}$ on top of the opaque-at-all-wavelengths 0.725 $R_{\rm J}$  
planet.  This large distance makes it immediately clear that we would  
only be sensitive to species in a large extended exosphere, such as  
that observed by Vidal-Madjar \etal (2003).  Therefore, only if K and  
Li had a large enough column abundance to remain optically thick past  
2.1 $R_{\rm J}$ would these atoms have been detected.

Furthermore, these species must remain neutral and atomic, such that  
they would absorb stellar flux at our specified wavelengths.  While  
these species are likely neutral in the lower atmosphere around $P  
\gtrsim$1 mbar, where Charbonneau \etal (2002) detected Na in HD  
209458b, all exospheric models for highly irradiated giant planets  
predict temperatures on the order of 10$^4$ K (Lammer \etal 2003,  
Yelle 2004, Tian \etal 2005).  Based on calculations by Yelle (2004)  
for an HD 2094358b-like planet at 0.04 AU, 2.1 $R_{\rm J}$ likely  
corresponds to $P \sim 10^{-10}$ bar and $T \sim 2 \times 10^4$ K.   
This high temperature makes it likely that Li and K will be ionized.   
Examining K specifically, Lodders \& Fegley (2006) review the  
chemistry of alkalis in substellar atmospheres and plot a curve where  
K I and K II have an equal abundance.  This curve has only weak  
pressure dependence.  At the lowest pressure considered, 1 mbar, K I  
and K II are equal in abundance at a temperature of 1900 K.  Our  
exospheric temperatures are likely 5-10 times hotter, meaning K II is  
strongly preferred.  Photodissociation is also likely important and  
will further reduce the abundances of Li I and K I.  Fortney \etal  
(2003) examined ionization of Na in the atmosphere of HD 209458b and  
found that photoionization of this species on the planet's limb could  
proceed to pressures as high as $\sim$ 1 mbar.  The atoms Li and K  
have even lower ionization potentials.  Due to the high temperatures  
and strong photoionizing flux, it appears unlikely that neutral  
atomic Li and K will be found in significant abundance in the  
extended atmosphere of HD 149026b.  This is consistent with the  
observations of Charbonneau \etal (2002), who observed Na I in the  
lower atmosphere, but not the exosphere, of HD 209458b.

R.~S.~Freedman has kindly provided us with absorption cross-sections  
as a function of wavelength for K and Li across our observation  
bands.  These are $\sim 4 \times 10^{-13}$ cm$^2$ for Li and $\sim 1  
\times 10^{-13}$ cm$^2$ for K.  Since we find that the exosphere has  
an optical depth $\lessim 1$ at 2.1 $R_{\rm J}$, this corresponds to  
column densities for these atoms in their neutral atomic ground  
state, of $\lesssim 2 \times 10^{12}$ cm$^{-2}$for Li, and $\lessim 9  
\times 10^{12}$ cm$^{-2}$ for K, at this radius.

A more sensitive search from
space-based telescopes could result in a detection of the planetary
atmosphere. Ionized alkali species would have noble gas electron 
configurations and would not 
have prominent absorption features even for space based observations.  We
advocate a search for H Lyman $\alpha$ from space to give a first  
estimate of the extent of an exosphere in HD 149026b.  At the same time, since  
there is a
clear link between heavy element mass and atmospheric metallicity in the
giant planets in our own solar system, the atmosphere and exosphere of
HD 149026b could be richer in metals than HD 209458b. The next elements
to search for would probably be C, N, and O species.

We gratefully acknowledge the dedication and support of the Keck
Observatory staff,
in particular Grant Hill for support with HIRES.  We thank Paul  
Butler and
Steve Vogt for contributing the observed spectra.  We thank Richard  
Freedman for providing us with opacity data.  Thanks to Jessica
Lovering for reduction of the CCD images to 1-D spectra.  We thank  
the NOAO
and NASA Telescope
assignment committees for generous allocations of telescope time.
Data presented herein were obtained at the W. M. Keck Observatory
from telescope time allocated to the National
Aeronautics and Space Administration through the agency's scientific
partnership with the California Institute of Technology and the
University of
California.  The Observatory was made possible by the generous financial
support of the W. M. Keck Foundation. We thank the Michaelson Science
Center for travel
support and support through the KDPA program.  DAF is a Cottrell
Science Scholar
of Research Corporation. We acknowledge support from NASA grant (to
DAF); a NASA Postdoctoral Program Fellowship (to JJF);
NASA grant NCC5-511; NASA grant NAG5-75005 (to GWM).

\clearpage

\begin{figure}
\begin{center}
\leavevmode
\scalebox{1}{\includegraphics{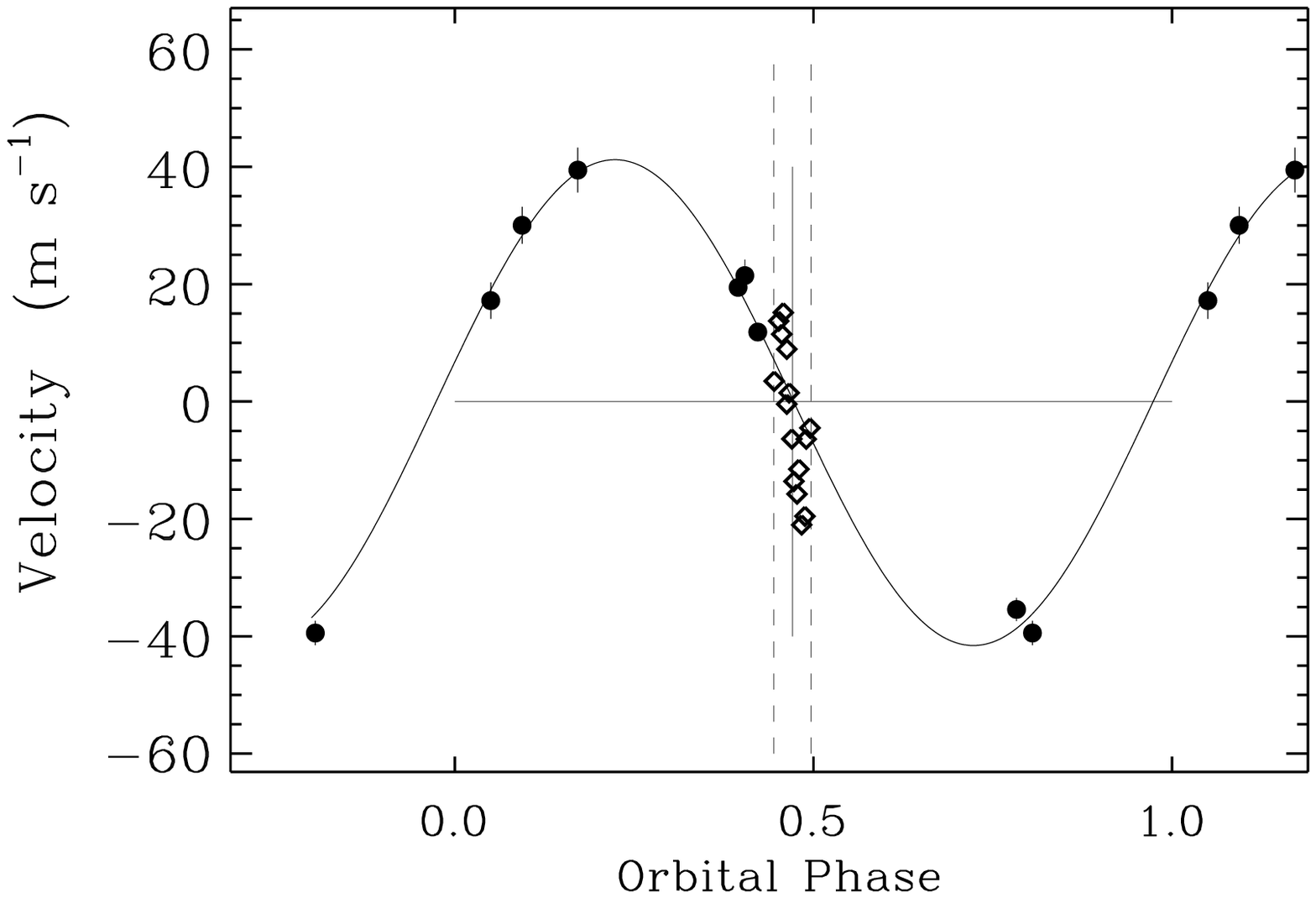}}
\end{center}
\caption{Updated phased radial velocities for HD 149026. The velocities
measured at Keck during transit are shown as open diamonds in the plot.
These measurements
exhibit the Rossiter-McLaughlin effect and were removed when fitting
a Keplerian model.\label{fig1}}
\end{figure}

\clearpage

\begin{figure}
\begin{center}
\leavevmode
\scalebox{0.75}{\includegraphics{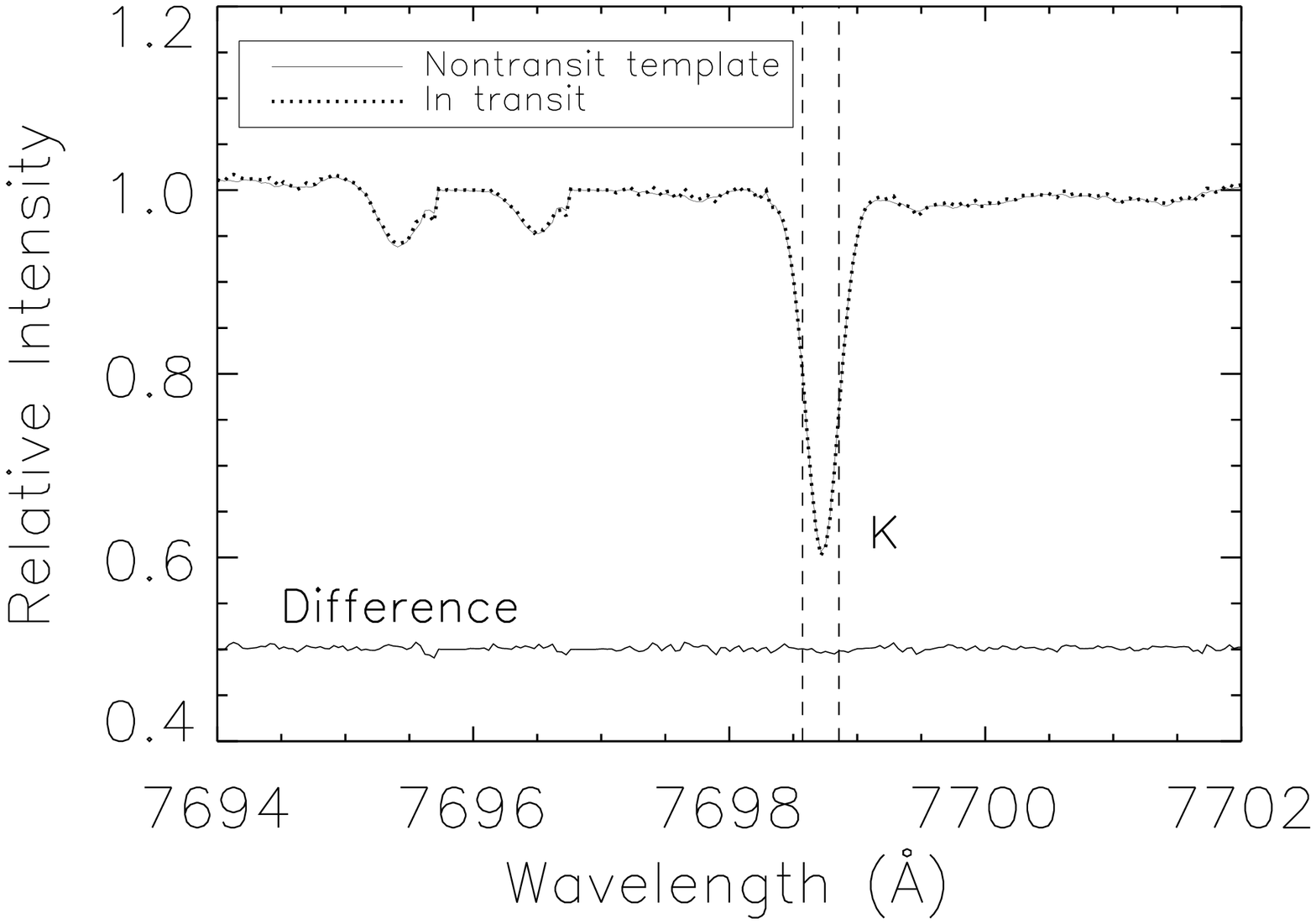}}
\end{center}
\caption{The template nontransit spectrum for the K line overplotted  
on a
transit spectrum. The difference of the two (offset from zero) and the
10.4 \kms wavelength bin (dashed lines) are also shown in
this plot.\label{fig2}}
\end{figure}

\clearpage

\begin{figure}
\begin{center}
\scalebox{1}{\includegraphics{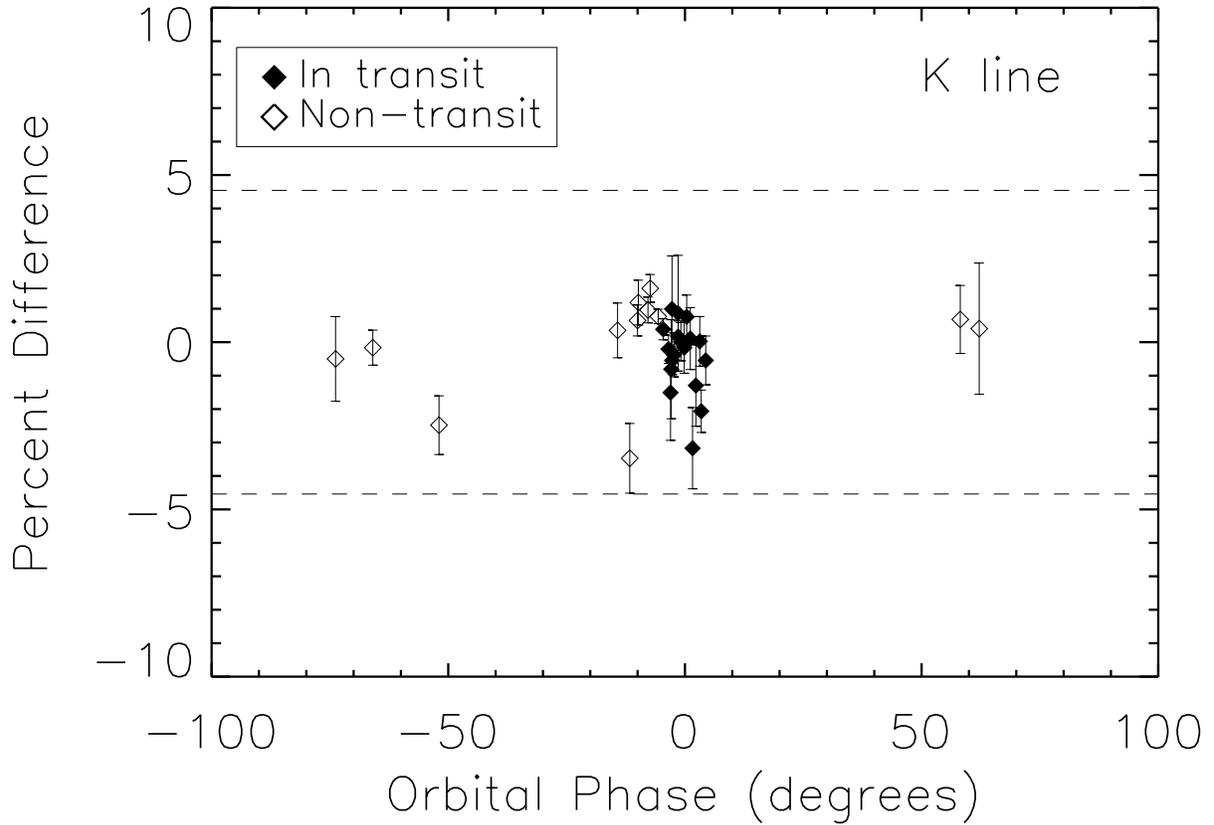}}
\end{center}
\caption{Deviations of the K line. The percent deviation is averaged  
over
the 10.4 \kms bin. This percent deviation was calculated by taking the
difference between each spectrum and the template nontransit spectrum
(which was constructed by averaging 12 nontransit spectra) and  
dividing by
the value of the template spectrum. The dashed lines indicate the
3$\sigma$ detection limits, where $\sigma$ is the RMS scatter in the
out-of-transit spectra.\label{fig3}}
\end{figure}

\clearpage

\begin{figure}
\begin{center}
\scalebox{0.75}{\includegraphics{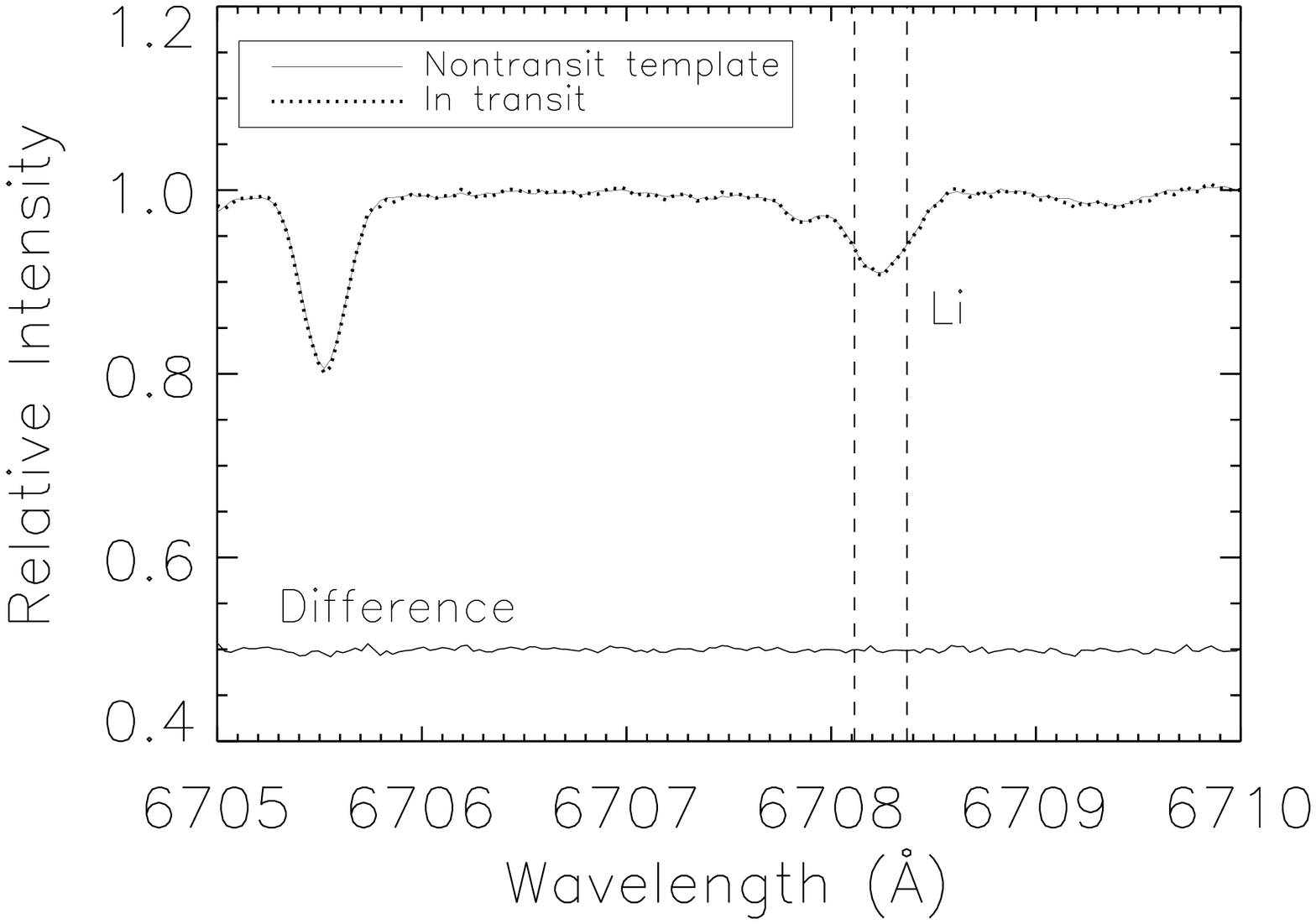}}
\end{center}
\caption{The template nontransit spectrum for the Li line overplotted  
on a
transit spectrum. The difference of the two (offset from zero) and the
10.4 \kms wavelength bin (dashed lines) are also shown in this
plot.\label{fig4}}
\end{figure}

\clearpage

\begin{figure}
\begin{center}
\scalebox{1}{\includegraphics{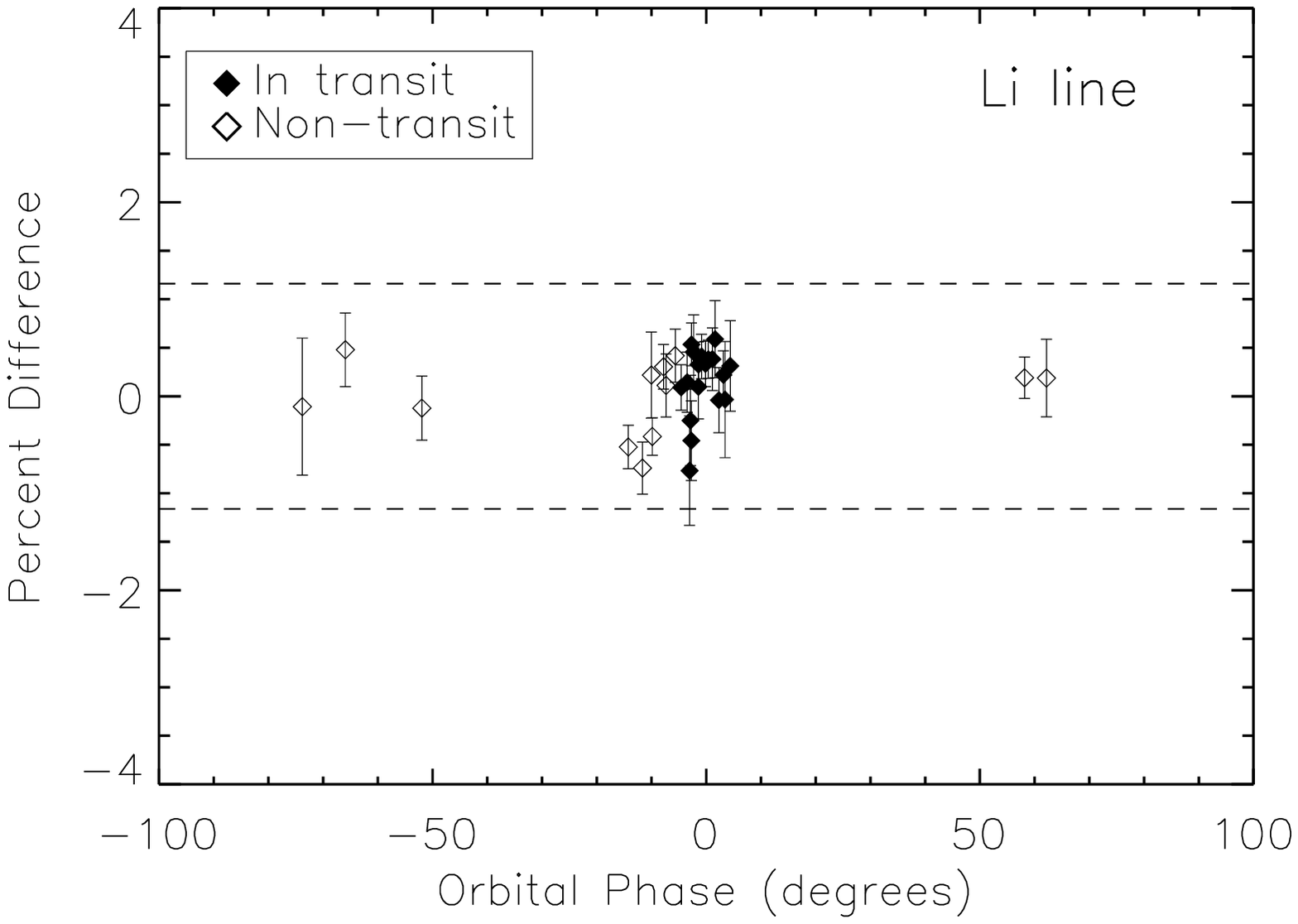}}
\end{center}
\caption{Deviations of the Li line. The percent deviation is averaged
over the 10.4 \kms bin. This percent deviation is calculated by  
taking the
difference between each spectrum and the template nontransit spectrum
(which is constructed by averaging 12 nontransit spectra) and  
dividing by
the value of the template spectrum. The 3$\sigma$ detection limits are
plotted as dashed lines.\label{fig5}}
\end{figure}

\clearpage

\begin{figure}
\begin{center}
\scalebox{1}{\includegraphics{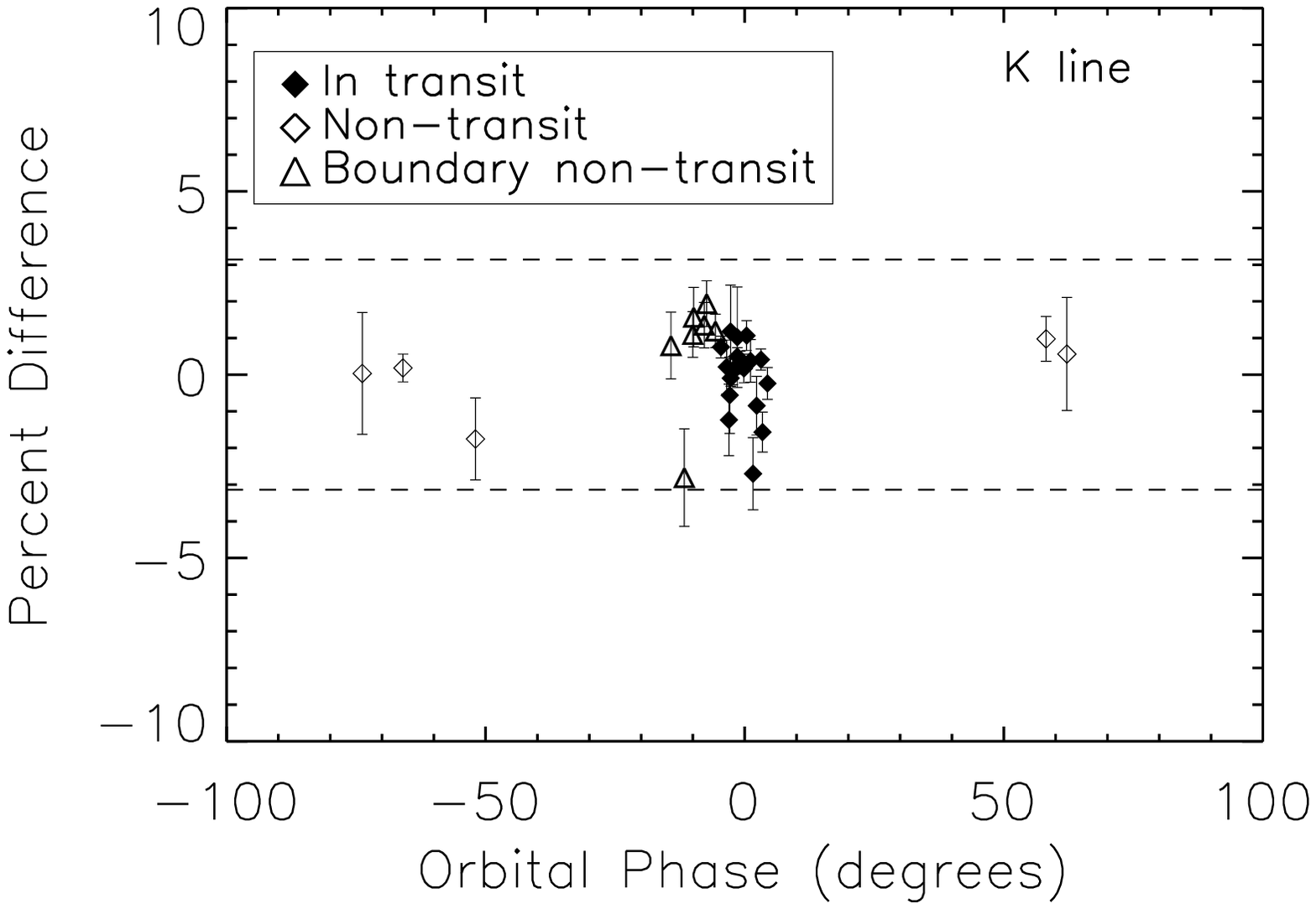}}
\end{center}
\caption{Deviations of the K line. These deviations were computed
using a template spectrum
that was constructed from 5 nontransit spectra. The triangles show
the boundary nontransit observations which were close to the ingress
and were not used in the template nontransit spectrum.\label{fig6}}
\end{figure}
\clearpage

\begin{figure}
\begin{center}
\scalebox{1}{\includegraphics{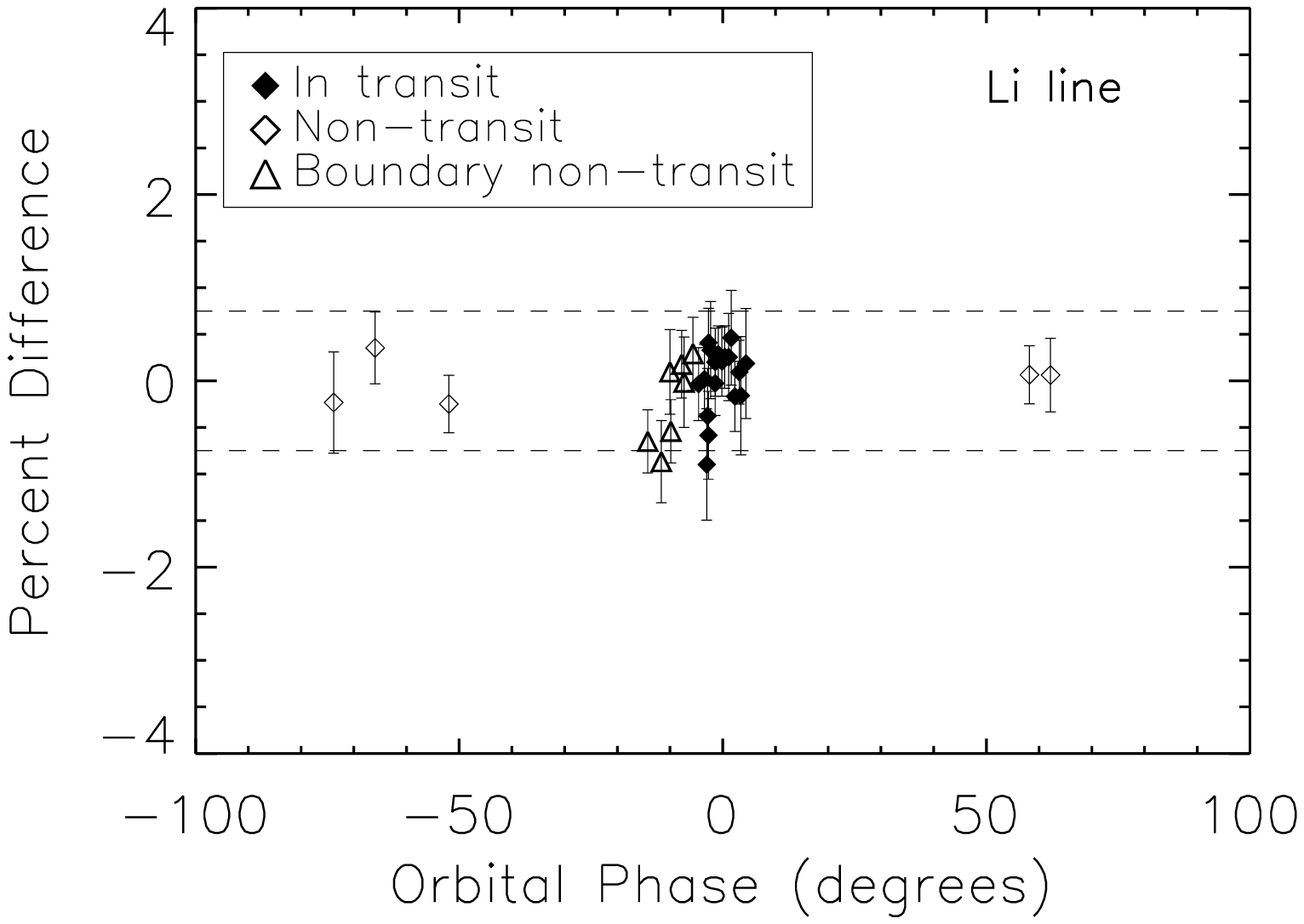}}
\end{center}
\caption{Deviations of the Li line. These deviations were computed
using a template
spectrum that was constructed from 5 nontransit spectra. The  
triangles show
the boundary nontransit observations which were close to the ingress
and were not used in the template nontransit spectrum.\label{fig7}}
\end{figure}

\clearpage

\begin{figure}
\centering
\scalebox{0.6}{\includegraphics{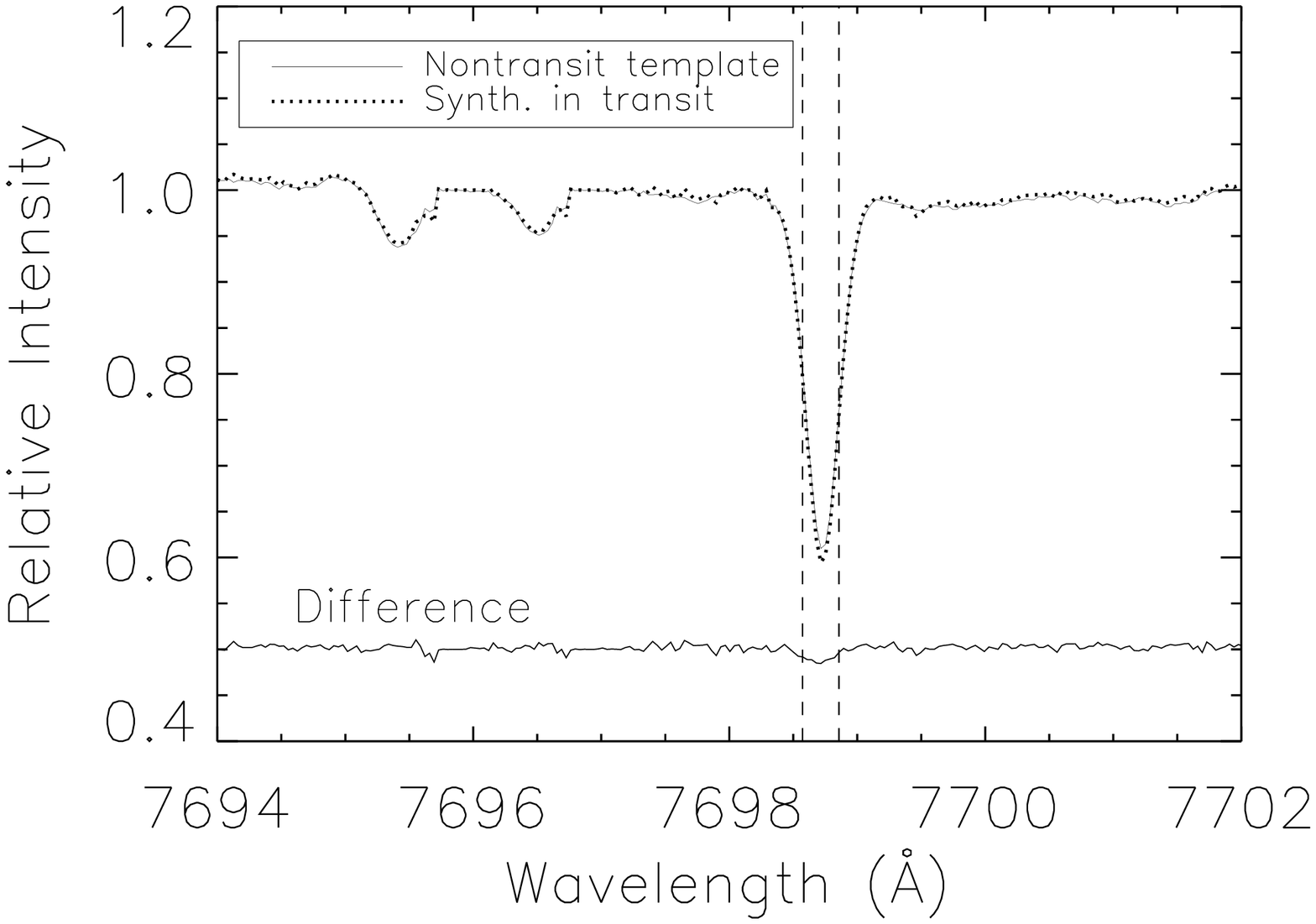}}
\scalebox{0.6}{\includegraphics{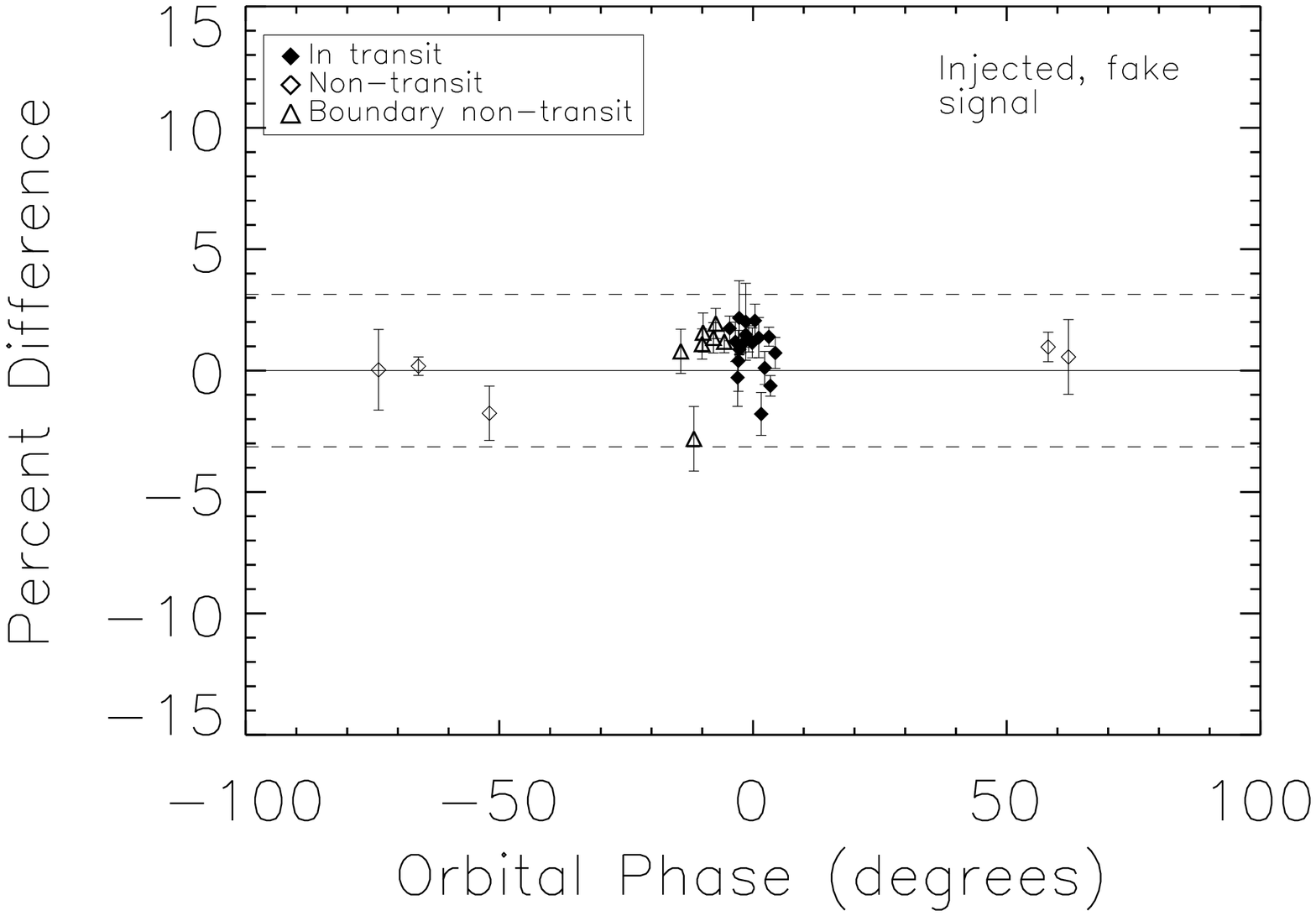}}
\caption{Synthetic data. A 2\% signal is introduced at 7698 \AA.
(a) The synthetic transit spectrum overplotted on
the template nontransit spectrum. The difference (offset from zero)
is also plotted. (b) The percent deviations
for such a signal. There is an 84\%\ confidence level to detect such a
signal.\label{fig8}}
\end{figure}

\clearpage

\begin{figure}
\centering
\scalebox{0.6}{\includegraphics{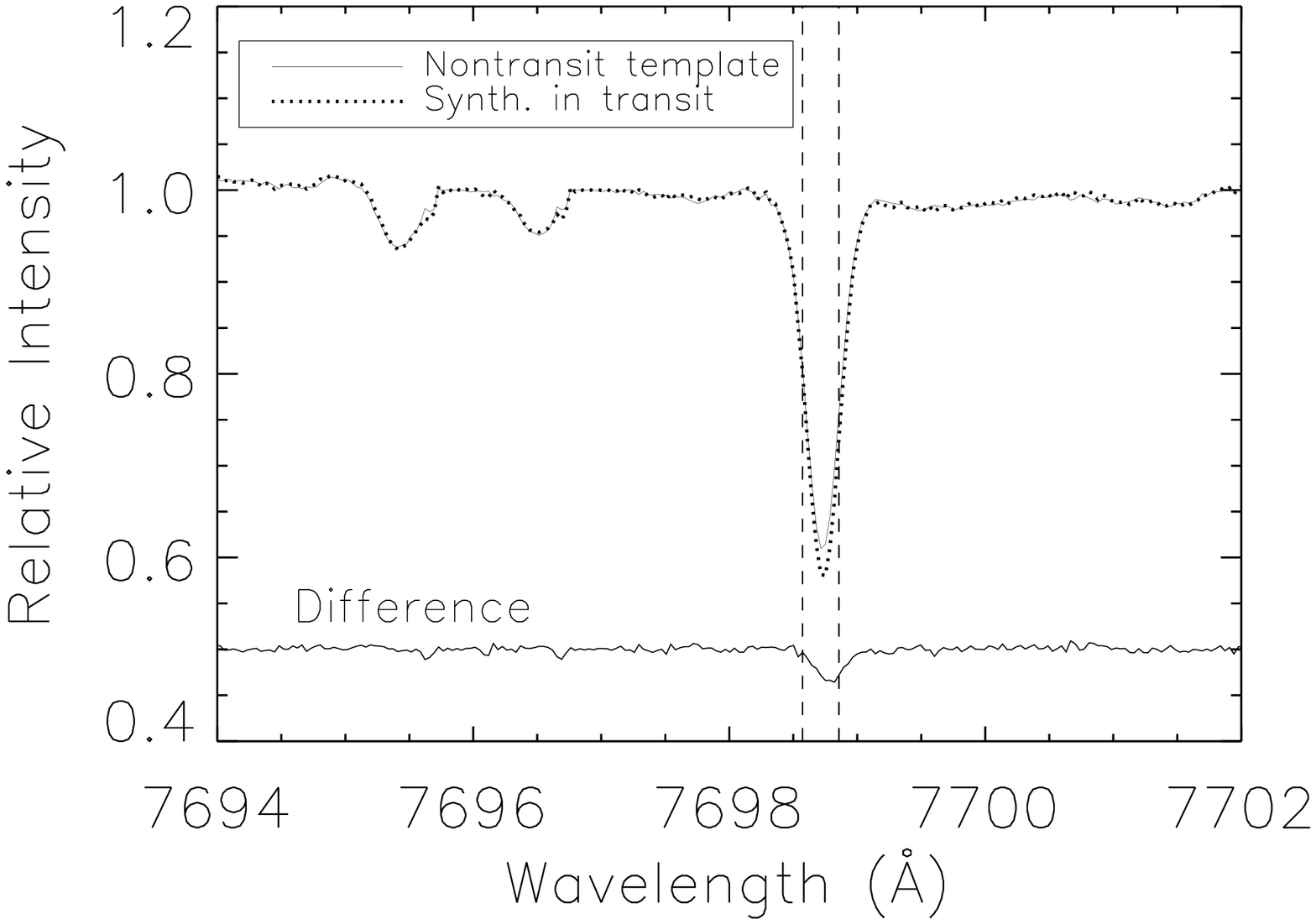}}
\scalebox{0.6}{\includegraphics{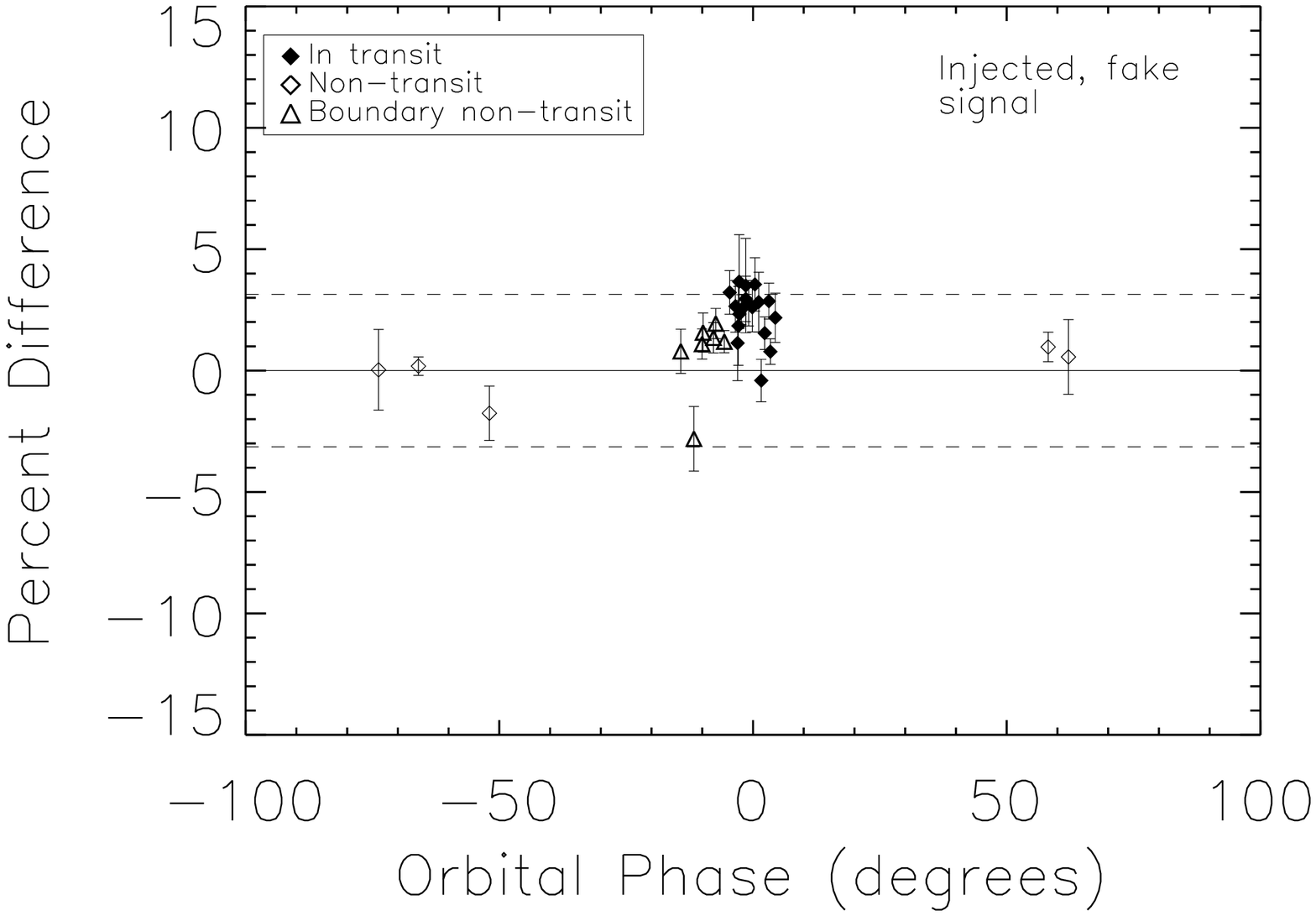}}
\caption{Synthetic data. A 5\% signal is introduced at 7698 \AA.
(a) The synthetic transit spectrum overplotted on
the template nontransit spectrum. The difference (offset from zero)
is also plotted. (b) The percent deviations
for such a signal. There is a 98.8 \%\ confidence level to detect such
a signal.\label{fig9}}
\end{figure}

\clearpage

\begin{figure}
\centering
\scalebox{0.6}{\includegraphics{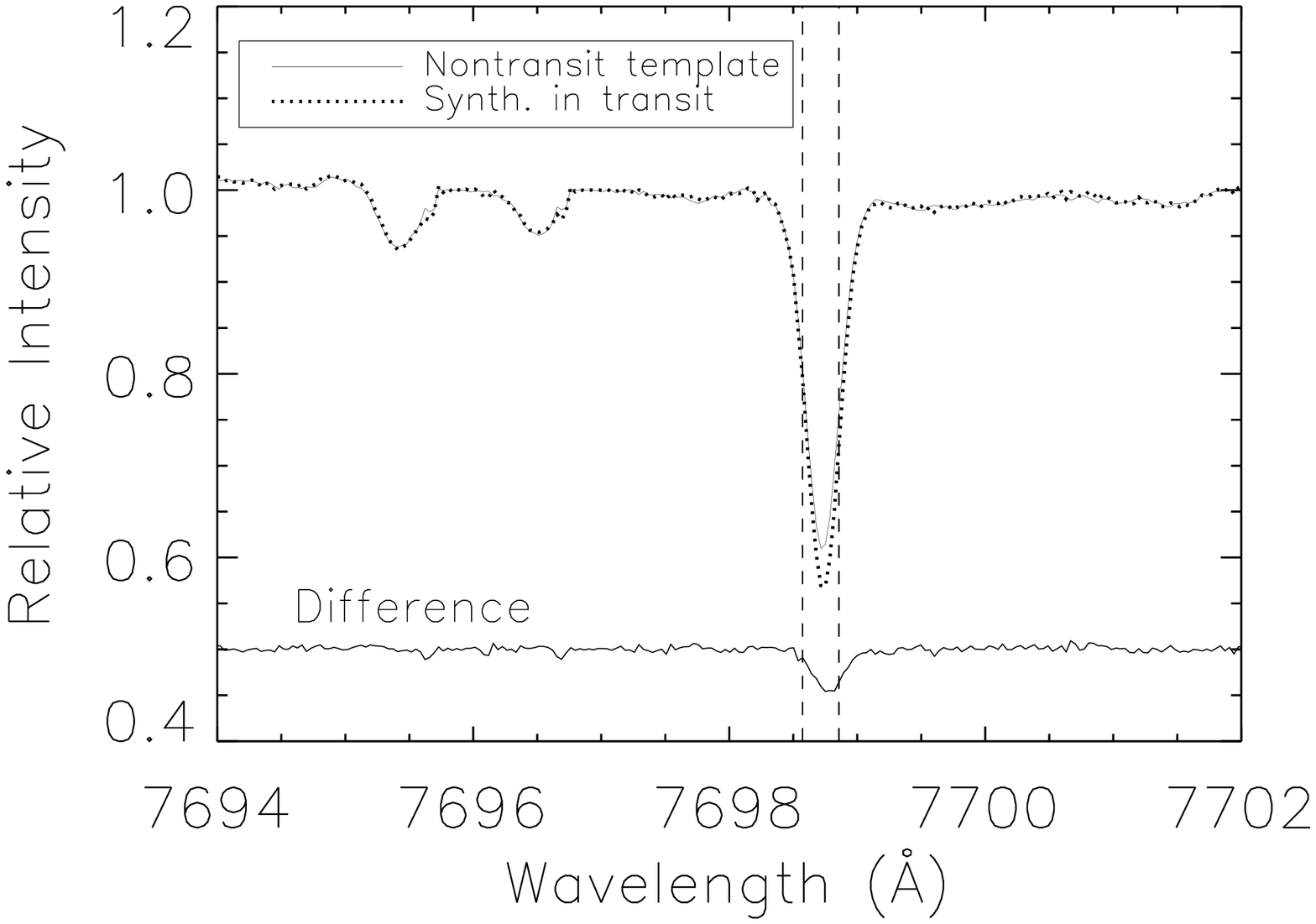}}
\scalebox{0.6}{\includegraphics{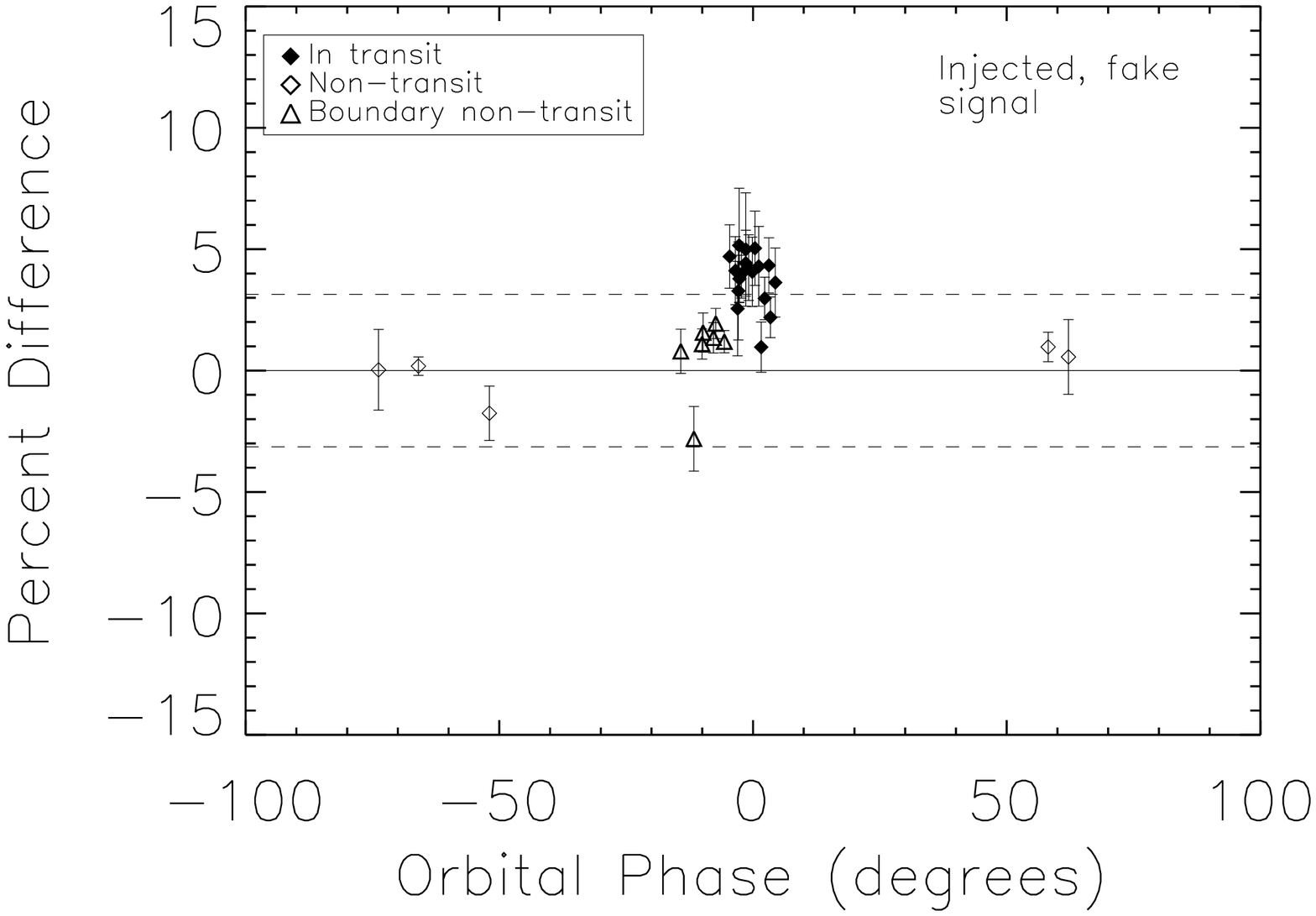}}
\caption{Synthetic data. An 8\% signal is introduced at 7698 \AA.
(a) The difference (offset from zero) of the synthetic transit
spectrum and the template nontransit spectrum. (b) The percent  
deviations
for such a signal. Such a signal would clearly be detected.\label 
{fig10}}
\end{figure}

\clearpage

\include{tab1.tex}
\include{tab2.tex}

\end{document}

%% file: tab1.tex
\begin{deluxetable}{ccc}
\tablecolumns{3}
\tablenum{1}
\tablewidth{0pt}
\tablecaption{K-S Test for Potassium}
\tablehead{\colhead{Pseudo Optical }&
\colhead{K-S }&
\colhead{Detection Confidence }& \\
\colhead{Depth Scaling Factor}&
\colhead{Significance Level }&
\colhead{Level (\%) } & \\ }
\startdata
1.00&0.9964&0.36\\
1.01&0.4131&58.69\\
1.02&0.1592&84.08\\
1.03&0.0484&95.16\\
1.04&0.0245&97.55\\
1.05&0.0117&98.83\\
1.08&0.0053&99.47\\
\enddata
\end{deluxetable}
\clearpage

%% file: tab2.tex
\begin{deluxetable}{ccc}
\tablecolumns{3}
\tablenum{2}
\tablewidth{0pt}
\tablecaption{K-S Test for Lithium}
\tablehead{\colhead{Pseudo Optical }&
\colhead{K-S }&
\colhead{Detection Confidence }& \\
\colhead{Depth Scaling Factor}&
\colhead{Significance Level }&
\colhead{Level (\%) } & \\ }
\startdata
1.00&0.6007&39.93\\
1.01&0.4131&58.69\\
1.02&0.1592&84.08\\
1.03&0.0484&95.16\\
1.06&0.0245&97.55\\
1.08&0.0117&98.83\\
\enddata
\end{deluxetable}
\clearpage

%% file: ms.bbl
\clearpage


\begin{thebibliography}

\bibitem[Brown (2001)]{brown01} Brown, T. M.  2001, \apj, 553, 1006

\bibitem[Bundy \& Marcy (2000)]{bm00}
Bundy, K. A., \& Marcy, G. W.  2000, PASP, 112, 1421

\bibitem[Burrows \& Sharp (1999)]{bs99}
Burrows, A., \& Sharp, C. M.  1999, \apj, 512, 843

\bibitem[Burrows \& Volobuyev (2002)]{bv02}
Burrows, A., \& Volobuyev, M.  2002, \apj, 583, 985

\bibitem[Butler et al. (1998)]{but98} Butler R. P., Marcy, G. W.,
Vogt, S. S., \& Apps, K.,  1998, PASP, 110, 1389

\bibitem[Charbonneau et al. (2002)]{char01}
Charbonneau, D., Brown, T. M., Noyes, R. W., \& Gilliland, R. L.   
2002, \apj, 586, 377

\bibitem[Charbonneau et al. (2005)]{char05}
Charbonneau, D., Winn, J. N., Latham, D. W., Bakos,
G., Falco, E. E., Holman, M. J., Noyes, R. W., \& Cs\'ak, B.
2005, \apj, in press

\bibitem[Deming \& Brown (2005)]{db05}
Deming, D., Brown, T. M., Charbonneau, D., Harrington, J., \&  
Richardson,
L. J. 2005, \apj, 622, 1149

\bibitem[Deming et al. (2005)]{dem05}
Deming, D., Seager, S., Richardson,
L. J., \& Harrington, J.  2005, Nature, 434, 740

\bibitem[Fischer et al. (2004)]{fisch04}
Fischer, D. A., Laughlin, G., Butler, P., Marcy, G.,
Johnson, J., Henry, G., Valenti, J., Vogt, S., Ammons, M.,
Robinson, S., Spear, G., Strader, J., Driscoll, P., Fuller, A.,
Johnson, T., Manrao, E., McCarthy, C., Munoz, M.,
Tah, K. L., Wright, J., Ida, S., Sato, B., Toyota, E.,
\& Minniti, D.  2004, \apj, 620, 481

\bibitem[Fischer et al. (2006)]{fisch06}
Fischer, D. A., Laughlin, G., Marcy, G. W., Butler, R. P.,
Vogt, S. S., Johnson, J. A.,
Henry, G. W., McCarthy, C., Ammons, M., Robinson, S.,
Strader, J., Valenti, J. A., McCullough, P. R.,
Charbonneau, D., Haislip, J., Knutson, H. A.,
Reichart, D. E., McGee, P., Monard, B., Wright, J. T.,
Ida, S., Sato, B., \& Minniti, D.  2006, \apj, 637, 1094

\bibitem[Fortney (2005)]{fort05}
Fortney, J. J. 2005, \mnras, 364, 649

\bibitem[Fortney (2003)]{fort03} Fortney, J. J., Sudarsky, D.,  
Hubeny, I., Cooper, C. S., Hubbard, W. B., Burrows, A., \& Lunine, J.  
I.  2003, \apj, 589, 615

\bibitem[Fortney et al. (2006)]{fortn06}
Fortney, J. J., Saumon, D., Marley, M. S., Lodders, K.,  \&
Freedman, R. S. 2005, \apj, 642, 495

\bibitem[Henry et al. (2000)]{henry00}
Henry, G. W., Marcy, G. W., Butler, R. P.,  \&
Vogt, S. S. 2000, \apj, 529, L41

\bibitem[Hubbard (2001)]{hubb01}  Hubbard, W. B., Fortney, J. J.,  
Lunine, J. I., Burrows, A., Sudarsky, D., \& Pinto, P.  2001, \apj,  
560, 413

\bibitem[Johnson (2006)]{john06}  Johnson, et al. 2006, in press

\bibitem[Kirkpatrick (2005)]{kirk05}  Kirkpatrick, J. D.  2005, ARA 
\&A, 43, 195

\bibitem[Lammer (2003)]{lamm03}  Lammer, H., Selsis, F., Ribas, I.,  
Guinan, E. F., Bauer, S. J., \& Weiss, W. W.  2003, \apj, 598, L121

\bibitem[Lodders (1999)]{lod99}  Lodders, K. 1999, \apj, 519, 793

\bibitem[Lodders and Fegley (2006)]{lod06}  Lodders, K., Fegley, B.,  
Astrophysics Update v. 2, in press, astro-ph/0601381

\bibitem[Sato et al.(2005)]{sat05} Sato, B., Fischer, D. A., Henry,
G. W., Laughlin, G., Butler, R. P., Marcy, G. W.,
Vogt, S. S., Bodenheimer, P., Ida, S., Toyota, E., Wolf, A.,
Valenti, J. A., Boyd, L. J., Johnson, J. A., Wright, J. T.,
Ammons, M., Robinson, S., Strader, J., McCarthy, C.,
Tah, K. L., \& Minniti, D.  2005, \apj, 633, 465

\bibitem[Seager \& Sasselov(2000)]{ss00}
Seager, S., \& Sasselov, D. D.  2000, \apj, 537, 916

\bibitem[Sudarsky et al. (2003)]{sud03} Sudarsky, D.,
Burrows, A., \& Hubeny, I.  2003, \apj, 588, 1121

\bibitem[Sudarsky et al. (2000)]{sud00} Sudarsky, D.,
Burrows, A., \& Pinto, P.  2000, \apj, 538, 885

\bibitem[Tian et al.(2005)]{tian05} Tian, F., Toon, O. B., Pavlov, A.  
A., \& De Sterck, H.  2005, \apj, 621, 1049

\bibitem[Vidal-Madjar et al.(2003)]{vm03}
Vidal-Madjar, A., Lecavelier des Etangs, A., D\'esert, J. M.,
  Ballester, G. E., Ferlet, R, H\'ebrard, G., Mayor, M.  2003, \nat,  
422, 143

\bibitem[Vogt(1994)]{vogt94} Vogt, S. S., et al.  1994, Proc.SPIE,  
2198, 362

\bibitem[Wolf(2006)]{wolf06} Wolf, A., et al.  2006, in press

\bibitem[Yelle(2004)]{yelle04} Yelle, R. V.  2004, Icarus, 170, 167

\end{thebibliography}
